\documentstyle[eqsecnum,aps]{revtex}
\begin{document}
\draft
\preprint{SU-GP-93/6-1, gr-qc/9306007}
\title{Pull Back Inariant Matter Couplings}
\author{Donald M. Marolf\cite{Marolf}}
\address{Physics Department, Syracuse University,
Syracuse, New York 13244} \date{June 3, 1993}
\maketitle

\begin{abstract}
An algorithm is described for the construction of actions for scalar,
spinor, and vector gauge fields that remains well-defined when the
metric is degenerate and that involve no contravariant tensor fields.
These actions produce the standard matter dynamics and coupling to gravity
when tetrad is nondegenerate, but have the property that all fields
that appear in them can be pulled back through an arbitrary map of degree one
and that this pull back leave the action invariant when the map has degree
one.
\end{abstract}

\section{Introduction}

One of the intriguing features of tetrad formulations of gravity is that
they remain well-defined when the tetrad becomes degenerate.  This
property has proved useful in the quantization of 2+1 gravity \cite{Witten}
and may be important for the quantization of 3+1 gravity as well (see
comments in \cite{Ash}).  The quantization of 2+1 gravity in fact used
the more startling property that nondegenerate tetrads are gauge
equivalent to certain degenerate tetrads.

A similar property has been discussed for the 3+1 case \cite{Gary}, where the
action is:
\begin{equation}
\label{first}
S = \int {\bf e}^a \wedge {\bf e}^b \wedge {\bf F}^{cd} \epsilon_{abcd}
\end{equation}
where ${\bf F}^{cd}$ is the field strength of either a spin connection
\cite{Palatini} or a self-dual connection \cite{Ash} and the bold-faced
print denotes differential forms, and ${\bf e}^a = e^a_{\mu} {\bf dx}^{\mu}$
is the tetrad one-form.  Horowitz points out that these fields may all be
pulled
back through an arbitrary smooth map $\phi$ and that the action is
invariant under this transformation if $\phi$ has degree one and uses
this observation to construct topology changing solutions to the equations of
motion.

However, this is valid only for the purely gravitational case
since the standard matter actions are {\em not} invariant under
pull backs and are not well-defined when the tetrad is degenerate.
The problem is that both the standard variational principles and the standard
equations of motion for the matter fields involve the inverse $e^{\mu}_m$ of
the
tetrad $e^m_{\mu}$ \footnote{We use greek letters for spacetime indices and
latin letters for internal Lorentz indices}.
For example, the usual scalar field action contains
two inverse tetrads:
\begin{equation}
S = \int d^4x \sqrt{g} g^{\mu \nu} \partial_{\mu}
\partial_{\nu} \phi = \int d^4x e \ e^{\mu}_a \eta^{ab}
e^{\nu}_b \partial_{\mu} \partial_{\nu} \phi \end{equation}

A suggestion was made in \cite{Gary} of how to avoid this problem, at least
if $e_a^{\mu}$ is degenerate only on a set of measure zero.  The idea is
to require
the fields to solve the equations of motion (Einstein's equations and
the matter equations of motion) only where $det(g) \neq 0$ but also
to be smooth everywhere, even when $det(g) = 0$.
This approach has been criticized on the grounds that it does
not follow from and may be in contradiction to a variational principle
for the system dynamics.
An example for which similar problems arise is described in \cite{cones} for
the pure gravity case.  There, a metric is defined on ${\cal R}^2 \times
D^2$ which, when
their parameter $\alpha$ is a negative integer,
is smooth everywhere and is non-degenerate
except at the center of the disk.  However, there is a conical singularity
in the spacetime at this center point.
The above prescription would allow an arbitrary
strength for this singularity (so long as $\alpha$ remains a negative
integer), but the Einstein-Hilbert variational
principle requires that the strength of this singularity vanish.

It is important to point out that it is {\em not} our goal here to
describe any dynamics of matter fields in the presence of a
degenerate metric that might follow from the standard variational
principles for matter fields by some regularization procedure.
In fact, we will not even ask that our
description be consistent with any such dynamics.  Instead, we take
the position that we should find {\em some} variational principle
that

\begin{enumerate}

\item [i)] Remains well-defined when the tetrad is degenerate

\item [ii)] Reproduces the usual dynamics for standard matter fields
whenever the tetrad is non-degenerate and

\item [iii)] Provides the proper source terms for the gravitational field.
\end{enumerate}

These need not, in general, require consistency with any dynamics
prescribed by the standard variational principles when the
tetrad is degenerate.  Again, an example of such a discrepancy is the
case of conical singularities in vacuum general relativity, for while
the metric given in \cite{cones}
is not a stationary point of the Einstein-Hilbert
action, the corresponding tetrad together with a connection that vanishes
{\em everywhere}, even at the singularity, {\em is} a stationary point of the
``first order action" \ref{first} with, say, the tetrad specified on the
boundary and the appropriate boundary terms included in the variational
principle.  This happens despite the fact that this variational principle
reproduces the Einstein-Hilbert dynamics when the tetrad is not
degenerate.

Actions for matter fields satisfying (i) - (iii)
have been discussed before \cite{others,others2},
but their description has always involved contravariant tensor fields.  For
example, the massless free scalar field can be described by the action:
\begin{equation}
S = \int  \sqrt{-g}
(\pi^{\mu} \partial_{\mu} \phi - \case{1}{2} \pi^{\mu} \pi^{\nu}
g_{\mu \nu}) \ d^4x
\end{equation}
Since such a formalism is not well-suited to the pull back construction of
\cite{Gary}, we would like to like to find action principles that
satisfy (i)-(iii) and contain only fields that can be pulled back through
an arbitrary smooth map.

By making use of differential forms as Lagrange multipliers, we will present
an algorithm in section \ref{alg} which transforms actions for standard
matter fields into action principles that
involve
{\em no} contravariant tensor fields or inverted tetrads, but that
are equivalent to the original actions when the tetrad is nondegenerate.
This algorithm can be applied to arbitrarily complicated systems, though
the fields involved will have to be of a certain type.  However, because
all scalar fields, all spinor fields,
and all vector gauge fields {\em are} of this type, this is not a severe
restriction.  We show in sections \ref{proof} and \ref{source} that these
actions have properties (ii) and (iii) above.
A similar
scheme exists for arbitrary fields, but now contravariant fields
must be introduced.  However, this method introduces fewer contravariant
fields that previous prescriptions
\cite{others,others2} and still allows a sort of pull back in certain cases.
We will concentrate on the cases in which the contravariant fields
can be completely eliminated and describe the more
general case in section \ref{gen}.  We close with a discussion of the
gauge invariance of the resulting actions.

\section{New Actions for Old}
\label{alg}

We will now present an algorithm that will construct
from an action $S^0$ an action $S$ such that

\begin{enumerate}

\item [i)]  $S$ leads to the same dynamics as $S^0$ when the tetrad is
non-degenerate.

\item [ii)] $S$ is well-defined for arbitrary smooth tetrads, regardless
of degeneracies.

\item [iii)] $S$ provides the same coupling to the gravitational
fields as $S^0$ when the equations of motion are satisfied.

\item [iv)] $S$ contains only {\em covariant} tensor fields and scalar
fields so that all fields in the action can be pulled back through an
arbitrary smooth map.  In fact, the
action $S$ is invariant under pull backs of degree one.

\end{enumerate}

Our algorithm will apply only to the limited class of actions in which
both the matter fields and their derivatives appear only in an
appropriately antisymmetrized
form, though we allow arbitrary combinations of tetrads and
inverse tetrads.
The restriction on matter fields
should come as no surprise after consideration of
the action \ref{first}, in which the use of differential forms eliminates
any need for the inverse tetrads or contravariant tensor fields.  The
antisymmetry condition guarantees that the matter fields can be replaced
by differential forms and that, by introducing additional differential
forms as Lagrange multipliers, the entire action can be written in terms
of differential forms.

There is one more intuitive idea that we should present before describing
the algorithm.  This idea was also inspired by \cite{Gary}, in which the
suggestion was made that solutions to the field equations be considered
for which scalars formed from the fields remain finite, but the individual
fields may diverge or go to zero.  For example, $e^{\mu}_a$ may diverge
but only if the derivatives of scalar fields $\partial_{\mu} \phi$
vanish fast enough that $e^{\mu}_a \partial_{\mu} \phi$ remains well-defined.
We note that the contractions $e^{\mu}_a \partial_{\mu} \phi \equiv \phi_{,a}$
are just the internal components of $d\phi$ and, in order to guarantee
that they take on finite values, we will take these internal components to
be the fundamental description of $d\phi$.

This brings us to the algorithm itself.  The description below refers to a
four-dimensional spacetime and uses the word ``tetrad" to refer to the fields
$e_{\mu}^a$.  However, for our purposes the dimensionality of the manifold is
completely unimportant and so is the explicit form of the gravitational
action so long as gravity is described by a ``tetrad" and a connection.
Thus, the procedure works equally well for 2+1, 3+1, or higher
dimensional systems and also for higher derivative theories.  We use
four-dimensional language only for convenience.

Suppose then that we are given some action $S^0$ that is an integral over
a four manifold ${\cal M}$ with boundary $\partial {\cal M}$ of a
four-form ${\cal L}^0$ which is a {\em function} of completely
antisymmetric covariant
matter fields $f^{(i)}_{[\mu_1 \mu_2 ...]}$ of density weight
zero and their
antisymmetrized covariant derivatives ${\cal D}_{[\alpha} f^{(i)}_{\mu_1 \mu_2
...]}$ as well as the tetrad $e^a_{\mu}$ and its inverse $e_a^{\mu}$.
This covariant derivative
is to be given by the Lorentz\footnote{By ``Lorentz" connection we mean
$SU(1,1)$, $SO(3,1$, $SL(2,C)$, or whatever gauge group is appropriate
for the dimension of spacetime and the description of gravity under
consideration.} connection
$\omega^a{}_{b \mu}$ that describes gravity and
acts only on internal indices\footnote{Nevertheless, we have just
stated that the covariant
derivative acts on objects $f^{(i)}_{[\mu_1 \mu_2 ...]}$
that have spacetime indices as well.  Because of the
antisymmetrized form  ${\cal D}_{[\alpha} f^{(i)}_{\mu_1 \mu_2 ...]}$
in which the covariant derivatives are assumed to appear,
the action will in
fact be independent of the extension of this covariant derivative to
act on such fields so long as it
is torsion-free.  For the purposes of our variational
principle, any fixed (i.e., field-independent) extension will do.}.
Note that only spacetime
indices on the fields $f^{(i)}$ have been indicated and that there is no
antisymmetry requirement on any internal indices that may be present
in the collective label.
We assume that appropriate boundary terms are also included in the action so
that functional derivatives are well-defined (with some set of boundary
conditions) but we do not keep track of such terms here.

In this case\footnote{Note
that this case includes arbitrary couplings of scalar and gauge fields!},
the following procedure produces an action $S$ that satisfies (i)-(iv)
above:

\begin{enumerate}

\item [Step 1)]  Insert sufficient inverse tetrads to
write all tensor fields in terms of their covariant
components:
\begin{equation}
f^{(i) \alpha}{}_{\mu \nu ...} \rightarrow e^{\alpha}_a \eta^{ab}
e^{\beta}_b f^{(i)}{}_{\beta \mu \nu ...}
\end{equation}

\item [Step 2)]  Insert enough tetrads to write all undifferentiated
tensor fields in terms of their tetrad components.
More specifically, for each tensor field $f^{(i)}{}_{[\mu \nu
...]}$ introduce a
set of scalar fields $f^{(i)}_{[m n ...]}$ with the same number
of internal indices as the rank of the original tensor
field.  Then perform the
substitution:
\begin{equation}
f^{(i)}{}_{[\mu \nu ...]} \rightarrow e^m_{\mu} e^n_{\nu} ...
f^{(i)}{}_{[mn ...]}
\end{equation}
for each {\em undifferentiated} field $f^{(i)}{}_{[\mu \nu ...]}$ in
${\cal L}^0$.

\item [Step 3)]  For each tensor field $f^{(i)}{}_{[\mu \nu ...]}$
present in the original action, introduce another collection of spacetime
scalars $f^{(i)}_{[mn ...,a]}$ labeled by one more internal
index than the rank of the tensor field.  This
comma is {\em only} a grouping symbol and does not denote
{\em any} kind of differentiation.  Now, introduce these
new fields into the action by using them to replace the
covariant derivatives of the fields $f^{(i)}{}_{[\mu \nu ...]}$
according to the rule:
\begin{equation}
{\cal D}_{[\alpha} f^{(i)}{}_{\mu \nu ...]} \rightarrow e^a_{\alpha} e^m_{\mu}
e^n_{\nu} f^{(i)}_{[mn ...,a]}
\end{equation}
Note that no derivatives remain in the
Lagrange density after this substitution has been performed.

\item [Step 4)]  Replace any spacetime Levi-Civita
densities with the corresponding internal symbols:
\begin{equation}
\epsilon_{\alpha \beta \gamma \delta} \rightarrow
\epsilon_{abcd}
e^a_{\alpha}e^b_{\beta}e^c_{\gamma}e^d_{\delta}
\ \text{and} \ \epsilon^{\alpha \beta \gamma \delta}
\rightarrow \epsilon^{abcd}
e_a^{\alpha}e_b^{\beta}e_c^{\gamma}e_d^{\delta}
\end{equation}

\item [Step 5)]  Formally cancel all contracted tetrads and
inverse tetrads:
\begin{equation}
e_m^{\mu} e_{\mu}^n \rightarrow \delta_m^n \ \text{and} \
e_{\mu}^m e_m^{\nu} \rightarrow \delta_{\mu}^{\nu}
\end{equation}
Note that since {\em all} matter fields and Levi-Civita
tensors have been replaced by fields with no spacetime
indices and since the connection and covariant derivative
no longer appear in the
Lagrange density after Step 5, any spacetime indices on tetrads
still present in the Lagrange density must be contracted with
spacetime indices from inverse tetrads and vice versa.
Because the Lagrange density is a four-form
and the action contains no matter densities, matter differential forms,
or covariant derivatives of tetrads,
after this step the tetrad
appears in the matter action only though the four-form
$\case {1}{4!} {\bf e}^a \wedge {\bf e}^b  \wedge {\bf e}^c
\wedge {\bf e}^d$.

\end{enumerate}

We now have a matter ``action functional" that depends
only on a set of scalar fields and the one-forms
${\bf e}^a$.  It is therefore perfectly well-defined when these
one-forms are degenerate and is also invariant under
pull backs.  However, since the tetrad only appears in the action though
the volume element $\case {1}{4!} {\bf e}^a \wedge {\bf e}^b  \wedge {\bf e}^c
\wedge {\bf e}^d$ this action produces the wrong coupling to the
gravitational field.  Even worse, it contains
no derivatives at all and so cannot lead to {\em any}
dynamics for the matter fields.  To correct these problems we
add one more step to our algorithm:

\begin{enumerate}

\item  [Step 6)]  For each field $f^{(i)}{}_{[\mu \nu ...]}$
with $n$ spacetime indices in the original action,
introduce a set of ($4-n$)-form fields
$\kappa^{(i)}$ with a number of internal indices
equal to the rank of the tensor field and a set of
(3-n)-form fields $\lambda^{(i)}$ \footnote{Of course, if $n > 4$,
the field does not appear in the action at all since its indices
must be completely antisymmetrized.  Similarly, if $n=4$, its derivatives
do not appear in the action in which case the $\lambda$ fields are not
needed and if $n=0$ the fields ${\kappa}$ are not needed.
The appropriate modifications can then be made to the following
discussion, but we will not treat this case explicitly and will
implicitly assume that $1 \leq n \leq 3$.} also with a
number of internal indices given by the rank of the
tensor field.  Now, use these new fields as Lagrange
multipliers and add to the above Lagrange density the
constraint terms:
\begin{equation}
\label{kappa}
\kappa^{(i)} \wedge  [f^{(i)}{}_{[\mu \nu ...]} {\bf dx}^{\mu} \wedge
{\bf dx}^{\nu} \wedge ... - f^{(i)}{}_{[mn ...]} {\bf e}^m \wedge {\bf e}^n
\wedge ...]
\end{equation}
and
\begin{equation}
\label{lambda}
\lambda^{(i)} \wedge [{\cal D} \wedge (f^{(i)}{}_{[\mu \nu ...]}
{\bf dx}^{\mu} \wedge {\bf dx}^{\nu} \wedge ...) - f^{(i)}{}_{[mn ...,a]}
{\bf e}^a \wedge {\bf e}^m \wedge {\bf e}^n \wedge ...]
\end{equation}
where the ${\cal D} \wedge$ with no subscript is the covariant exterior
derivative operator defined by ${\cal D}$.
\end{enumerate}

Intuitively, these constraints link the tetrad components
of the matter fields to the new fields that we have
introduced.  Practically, these
constraints reintroduce dynamics through the derivatives
in Eq. \ref{lambda} and reintroduce the proper coupling of
the matter to the gravitational fields through the tetrads they contain.
We will demonstrate these
practical properties in the next two sections, but first we note that when
this algorithm is applied to the massive scalar and Yang-Mills
actions, the results are:
\begin{equation}
S_{\phi} = \int (\case{1}{2} \phi_a \phi^a + m^2 \phi^2) \case{1}{4!}
\epsilon_{ijkl} {\bf e}^i
\wedge {\bf e}^j \wedge {\bf e}^k \wedge {\bf e}^l + \int \lambda \wedge
({\cal D} \phi - \phi_a {\bf e}^a)
\end{equation}
\begin{eqnarray}
S_{YM} &=& - \case{1}{4} \int (A_{[a,b]}^i - C^i_{jk} A_a^j A_b^k)
(A_{[c,d]}^i - C^n_{ml} A_c^m A_d^l) \eta^{ac} \eta^{bd}
 \case{1}{4!}
\epsilon_{a_1 a_2 a_3 a_4} {\bf e}^{a_1}
\wedge {\bf e}^{a_2} \wedge {\bf e}^{a_3} \wedge {\bf e}^{a_4} \cr
&+& \int \kappa_i \wedge (A^i_{\mu} {\bf dx}^{\mu} - A^i_a {\bf e}^a)
\cr &+& \int \lambda_i \wedge ({\cal D} \wedge A^i_\mu {\bf dx}^{\mu}
- A^i_[b,a] {\bf e}^{a} \wedge {\bf e}^b)
\end{eqnarray}
where in this last example $i,j,k,l,m,n$ are Yang-Mills indices, $\mu$ is
a spacetime index, and $a,b,c,d$ are internal Lorentz indices.

Note that our algorithm can easily be generalized to matter
Lagrangians that involve spinors.  To do so, we need only
replace each tetrad $e^a_{\mu}$ with a soldering form $\sigma_{\mu}^{AA'}$
in the appropriate steps above.  In fact, the algorithm treats Dirac
spinors much like scalar fields since the spinors have no
spacetime indices but only internal indices.

\section{Equivalence of New and Old Actions}
\label{proof}

There are several senses in which we might wish our new actions to be
equivalent to the original actions. Note, for example,
that the new and old
actions are numerically equal when the constraints hold,
as the constraints simply keep track of the various
substitutions made in the early steps of the
algorithm.  Here, however, we show that the matter dynamics
produced by the two actions are equivalent, or, more specifically,
that the matter
equations of motion that follow from
our new action are equivalent to the old matter equations of
motion {\em when the tetrad is non-degenerate}.  The presence of
spinor fields does not alter the discussion below.

Our new action $S$ is the integral of some four-form ${\cal L}$ obtained
from the original action $S^0 = \int {\cal L}^0 = \int L^0 \case{1}{4!}
\epsilon_{ijkl} {\bf e}^i \wedge {\bf e}^j \wedge {\bf e}^k \wedge
{\bf e}^l$
by following Steps 1-6.  Note that our fields $f^{(i)}{}_{[mn...]}$ appear
in ${\cal L}$ only as a result of Step 2 so that the
variation of $S$ with respect to these fields can be
expressed as
\begin{eqnarray}
\label{varfields}
0 &=& \case{1}{4!} \epsilon_{\alpha \beta \gamma \delta} {\bf dx}^{\alpha}
\wedge {\bf dx}^{\beta} \wedge {\bf dx}^{\gamma} \wedge {\bf dx}^{\delta}
{{\delta S} \over {\delta f^{(i)}{}_{[mn...]}}} \cr  &=&
{{\partial {\cal L}^0} \over {\partial f^{(i)}{}_{[\mu \nu
...]}}} {{\partial f^{(i)}{}_{[\mu \nu ...]}} \over {\partial
f^{(i)}{}_{[mn...]}}}  - \kappa^{(i)} \wedge {\bf e}^m \wedge {\bf e}^n ...
\end{eqnarray}
The substitutions of steps 1-5 are to be
performed after the derivative $\partial {\cal L}^0 / \partial f$
has been computed.  Note
that this derivative has a number of free spacetime indices
which can only appear through inverse tetrads after the
substitutions have taken place.  Thus,
\begin{equation}
\label{varf}
{{\partial
{\cal L}^0} \over {\partial f^{(i)}{}_{[\mu \nu ...]}}} =
D^{(i)mn...} e_m^{\mu} e_n^{\nu} ...
\case{1}{4!} \epsilon_{ijkl}
{\bf e}^i \wedge {\bf e}^j \wedge {\bf e}^k \wedge {\bf e}^l
\end{equation}
for some internal tensor field $D^{(i)mn.....}$ built entirely from the
matter fields.  The
partial derivative of $f^{(i)}{}_{[\mu \nu ...]}$ with respect to
$f^{(i)}{}_{[mn...]}$ is to be computed from Step 3, that is:
\begin{equation}
\label{partfs}
{{\partial f^{(i)}{}_{[\mu \nu ...]}} \over
{\partial f^{(i)}{}_{[mn...]}}} \equiv e^m_{\mu} e^n_{\nu}
... \end{equation}
The tetrads in \ref{varf} then cancel with the inverse tetrads in \ref{partfs}.
Strictly speaking, since expression \ref{varf} involves inverse
tetrads, it is not defined when the tetrad is
degenerate.  In that case, it is only our notation ${{\partial {\cal L}^0}
\over {\partial f}}$ and ${{\partial f} \over {\partial f}}$ that is
ill-defined
and the functional derivative \ref{varfields} is still perfectly well-defined
and equal to $D^{(i)}_{mn....} det(e)$.  However, we are interested
here in the case where the tetrad is {\em not} degenerate
so that there is no ambiguity in either
expression \ref{varf} or \ref{partfs}.  Despite this complication, we will find
this notation to be convenient.

Similarly, we will be interested in the variation of the
new action with respect to the ``derivative fields"
$f^{(i)}{}_{[mn...,a]}$:
\begin{equation}
\label{varderivatives}
0 = {{\delta S} \over {\delta f^{(i)}{}_{[mn...,a]}}} =
{{\partial {\cal L}^0} \over {\partial {\cal D}_{[\alpha}
f^{(i)}{}_{\mu \nu ...]}}} {{\partial {\cal D}_{[\alpha} f^{(i)}{}_{
\mu \nu ...]}} \over {\partial f^{(i)}{}_{[mn...,a]}}}  -
\lambda^{(i)} \wedge {\bf e}^a \wedge {\bf e}^m \wedge {\bf e}^m
\end{equation}
Here, the notation is similar to that used in Eq.
\ref{varfields} and the partial derivative ${{\partial {\cal D} f} \over
{\partial f}}$ is to be
computed using Step 3.

In order to show that the dynamics generated by these
equations is the same as what follows from the original
equations of motion ${{\delta S^0} \over {\delta f^{(i)}_
{[\mu \nu ...]}}} = 0$, we will, of course, need to relate
the new fields $f^{(i)}{}_{[mn....]}$ to the old fields
$f^{(i)}{}_{[\mu \nu ...]}$.  The proper relationship is
guaranteed by the constraints:

\begin{eqnarray}
\label{varlambda}
0 &=& {\bf V}
{{\delta S} \over {\delta \lambda^{(i)}{}_{[\alpha \beta ...]}}} =
{\bf dx}^{\alpha} \wedge {\bf dx}^{\beta} \wedge ...\{
{\cal D} \wedge (f^{(i)}_{[\mu \nu ...]}
{\bf dx}^{\mu} \wedge  {\bf dx}^{\nu} \wedge  ...) \cr
&-& f^{(i)}{}_{[mn....,a]}
{\bf e}^a \wedge {\bf e}^m \wedge {\bf e}^n \wedge ... \}
\end{eqnarray}

\begin{equation}
\label{varkappa}
0 = {\bf V}
{{\delta S} \over {\delta \kappa^{(i)}{}_{[\alpha \beta...]}}} =
{\bf dx}^{\alpha} \wedge {\bf dx}^{\beta} \wedge ... \{
f^{(i)}{}_{[\mu \nu ...]}
{\bf dx}^{\mu} \wedge {\bf dx}^{\nu} \wedge ... - f^{(i)}{}_{[mn...]} {\bf e}^m
\wedge {\bf e}^n \wedge ... \}
\end{equation}
and
\begin{equation}
\label{f}
0 = {\bf V} {{\delta S} \over {\delta f^{(i)}{}_{[\mu \nu...]}}} = \kappa^{(i)}
\wedge {\bf dx}^{\mu} \wedge {\bf dx}^{\nu} \wedge ... -
{\cal D} \wedge \lambda^{(i)}
\wedge {\bf dx}^{\mu} \wedge {\bf dx}^{\nu} \wedge ...
\end{equation}
where $\kappa^{(i)} = \case{1}{n!} \kappa^{(i)}{}_{[\alpha \beta ...]}
{\bf dx}^{\alpha} \wedge {\bf dx}^{\beta} \wedge ...$ and
$\lambda^{(i)} = \case{1}{m!} \lambda^{(i)}{}_{\alpha \beta ...}
{\bf dx}^{\alpha} \wedge {\bf dx}^{\beta} \wedge ...$, $n,m$ are
the rank of these forms, and we have introduce ${\bf V} =
 \case{1}{4!} \epsilon_{\alpha \beta \gamma \delta}
{\bf dx}^{\alpha} \wedge {\bf dx}^{\beta} \wedge {\bf dx}^{\gamma} \wedge
{\bf dx}^{\delta}$ to simplify the notation.
Note that equations \ref{lambda} and \ref{f}
contain the only derivatives in any of the
equations of motion.  We will see that the second of these effectively
contains all the dynamics of
the theory.

Our goal now is to show that when the tetrad is non-degenerate
the above equations of motion are equivalent to
the original equations of motion together with the following {\em definitions}:
\begin{mathletters}
\label{defs}
\begin{equation}
f^{(i)}{}_{[mn ...]} \equiv e_m^{\mu}
e_n^{\nu} f^{(i)}{}_{[\mu \nu  ...]}
\end{equation}
\begin{equation}
f^{(i)}{}_{[mn ...,a]} \equiv
{\cal D}_{[\alpha}f^{(i)}{}_{[\mu \nu ...]} e_a^{\alpha} e_m^{\mu}
e_n^{\nu}
\end{equation}
\begin{equation}
(*\kappa^{(i)})_{[\mu \nu ...]} \equiv
{{\partial L^0} \over {\partial f^{(i)[\mu \nu
...}}}
...
\end{equation}
\begin{equation}
\label{deflambda}
\lambda^{(i)}_{mn...} \equiv
* \bigl( {{\partial {\cal L}^0} \over {\partial {\cal D}^{\alpha}
f^{(i) [\mu \nu ...]}}} {\bf dx}^{\mu} \wedge {\bf dx}^{\nu} \wedge
... \bigr)
\end{equation}
\end{mathletters}
where $*$ is the Hodge duality operator.
We note that $*$ is well-defined and that
such definitions are always possible when the tetrad is
nondegenerate.

We also note that these are exactly the solutions of equations
\ref{varfields}, \ref{varderivatives}, \ref{varlambda}, and \ref{varkappa} when
the tetrad is nondegenerate.  Therefore, if the two theories are equivalent,
all of the dynamics must be contained in the single unsolved equation
\ref{f}.  This follows since direct substitution of
the results/definitions \ref{defs}
into Eq. \ref{f}, gives, after taking a dual:
\begin{equation}
0 = {{\partial {\cal L}^0} \over {\partial f^{(i)}{}^{[\mu \nu ... ]}}}
{\bf dx}^{\mu} \wedge {\bf dx}^{\nu} \wedge ... - * \Biggl( {\cal D} \wedge
* \biggl( {{\partial {\cal L}^0} \over {\partial ({\cal D}^{\beta}
f^{(i) [\lambda \kappa ... ]}}) } {\bf dx}^{\lambda} \wedge {\bf dx}^{\kappa}
\wedge ... \biggr)
\Biggr)
\end{equation}
which is just the set of matter equations of motion of our original action
$S^0$.

\section{Source Terms}
\label{source}

If the new actions are to be satisfactory they must produce not only
the correct matter field dynamics when the tetrad is nondegenerate,
but also the correct coupling to the gravitational fields
-- both to the connection and to the tetrad.  Specifically, we show
in this section that the gravitational source terms are identical for the old
and new actions when the tetrad is non-degenerate and the matter
equations of motion hold.
Again, no changes are needed to include spinors other than the
substitution of soldering forms for tetrads.
Because it is somewhat simpler, we consider the coupling to
the connection first.

We assume that our
description of gravity is based on a connection $\omega_{\mu}{}^{a}{}_b$ where
$\mu$ is a one-form index and $a,b$ are matrix indices in the vector
representation of the gauge group.  The covariant
derivative then acts on an internal vector by: ${\cal D}_{\alpha}v^a =
{\partial}_{\alpha} v^a + \omega_{\alpha b}{}^a v^b$.  Note that
${\cal D}_{[\alpha} f^{(i)}_{\mu \nu ...]} =
{\partial}_{[\alpha} f^{(i)}_{\mu \nu ...]}$ unless $f^{(i)}$ has internal
indices so that it will be important to display such indices explicitly.
We will write $f^{(i)}{}_{[\mu \nu ...]}
= f^{(i)}{}_{[\mu \nu ...]}|_{abc...}$ and $f^{(i)}{}_{[mn ...]}
= f^{(i)}{}_{[mn ...]}|_{ab...}$ where the extra internal indices are
displayed after the bar in order to separate them from internal
indices created by our algorithm that replace spacetime
indices on the original fields.  Note that since these extra indices were
previously absorbed into the collective label $(i)$, they should also
be displayed on the lagrange multipliers.  In particular, $\lambda^{(i)}
= \lambda^{(i)}|_{abc...}$.

The source term for the connection given by our new action is:
\begin{equation}
\label{connection}
{\bf V} {{\delta S} \over {\delta \omega_{\sigma}{}^a{}_b}} =
\sum_{n,i} \lambda^{(i)}|^{[a_1 a_2 ... a_{n-1} b
a_{n+1} ...]} f^{(i)}{}_{[\mu \nu ...]}|_{[a_1 a_2 ... a_{n_1} a
a_{n+1} ...]} \wedge {\bf dx}^{\sigma} \wedge {\bf dx}^{\mu}
\wedge {\bf dx}^{\nu} \wedge ...
\end{equation}
since the connection now appears in the action only through the
covariant derivative in the constraints.  However, we know that when
the tetrad is nondegenerate, $\lambda^{(i)}|^{ab...}$ is given
by Eq. \ref{deflambda} so that we have
\begin{eqnarray}
{\bf V}
{{\delta S} \over {\delta \omega_{\sigma}{}^a{}_b}} &=& \sum_{n,i}
* \biggl( {{\partial {\cal L}^0} \over {\partial {\cal D}_{[\alpha}
f^{(i)\mu_1 \mu_2 ...]}|{a_1 a_2 ... a_{n-1} a a_{n+1} ...}}}
{\bf dx}^{\alpha} \wedge {\bf dx}^{\mu_1} \wedge {\bf dx}^{\mu_2} \wedge
... \biggr) \cr &\wedge&
f^{(i)}{}_{[\nu_1 \nu_2 ...]}|^{a_1 a_2 ... a_{n-1} b a_{n+1} ...}
{\bf dx}^{\nu_1} \wedge {\bf dx}^{\nu_2} ... \cr
&\times& e_{\alpha}^c \epsilon_{ijkc}
{\bf e}^i \wedge {\bf e}^j \wedge {\bf e}^k \cr
&=& \sum_{i}
{{\partial {\cal L}^0} \over {\partial {\cal D}_{[\alpha}
f^{(i)}{}_{\mu_1 \mu_2 ...]}|_{a_1 a_2...}}}
{{\partial {\cal D}_{[\alpha}
f^{(i)}_{\mu_1 \mu_2 ...]}|_{a_1 a_2 ...}}
\over {\partial \omega_{\mu}{}^a_b}} \case{1}{4!}
\epsilon_{ijkl} {\bf e}^i \wedge
{\bf e}^j \wedge {\bf e}^k \wedge {\bf e}^l \cr
&=& {{\delta S^0} \over {\delta \omega_{\mu b}^a}}
{\bf V}
\end{eqnarray}
and in fact the source terms for the connection are equivalent in the two
actions.

This leaves only the variation with respect to the tetrad.  We note
that tetrad source terms from the new action can only arise from variation of
the volume element $ \case{1}{4!}
\epsilon_{ijkl} {\bf e}^i \wedge
{\bf e}^j \wedge {\bf e}^k \wedge {\bf e}^l$ that still appears in the
Lagrangian or from variation of the
constraints.  From our
new action then:

\begin{eqnarray}
\label{esource}
{\bf V}
{{\delta S} \over {\delta e^a_{\alpha}}} &=& \case{1}{3!}
 \epsilon_{abcd} {\bf dx}^{\alpha}
\wedge {\bf e}^b \wedge {\bf e}^c \wedge {\bf e}^d L^0 \cr
&-& \sum_{n,i} \bigl[f^{(i)}{}_{[m_1 m_2 ... m_{n-1}
a m_{n+1}
...]} \kappa^{(i)} \wedge {\bf e}^{m_1} \wedge {\bf e}^{m_2}
\wedge ... \wedge {\bf e}^{m_{n-1}} \wedge {\bf dx}^{\alpha} \wedge
{\bf e}^{m_{n+1}} \wedge ... \bigr] \cr
&-& \sum_{n,i}  \bigl[f^{(i)}{}_{[m_1 m_2 ... m_{n-1}
a m_{n+1}
..., b]} \lambda^{(i)} \wedge {\bf e}^b \wedge
{\bf e}^{m_1} \wedge {\bf e}^{m_2}
\wedge ... \wedge {\bf e}^{m_{n-1}} \wedge {\bf dx}^{\alpha} \wedge
{\bf e}^{m_{n+1}} \wedge ... \bigr] \cr
\cr &-&
\sum_i f^{(i)}{}_{[m_1 m_2 ...,a]}
\lambda^{(i)} \wedge {\bf dx}^{\alpha}
\wedge {\bf e}^{m_1} \wedge {\bf e}^{m_2} ...
\end{eqnarray}
where again the substitutions of Steps 1-5 are to be made in $L^0$ and
we have once again absorbed any internal indices on the original fields
into the collective label $(i)$.

Since none of our matter fields are densities, the original Lagrange density
${\cal L}^0$ must be of the form ${\cal L}^0 = L^0 \case{1}{4}
\epsilon_{ijkl} {\bf e}^i \wedge {\bf e}^j \wedge {\bf e}^k \wedge {\bf e}^l$
where all of the tetrads in $L^0$ appear contracted with some
$f^{(i)}{}_{[\mu \nu .....]}$ in the form $e^{\mu}_a f^{(i)}_{[\mu \nu
.....]}$,
$e^{\mu}_a
{\cal D}_{\alpha} f^{(i)}{}_{[\mu \nu .....]}$, or $e_a^{\alpha}
{\cal D}_{\alpha} f^{(i)}{}_{[\mu \nu .....]}$ and every spacetime index
associated with
either $f^{(i)}{}_{[\mu \nu .....]}$ or ${\cal D}_{\alpha}
f^{(i)}{}_{[\mu \nu .....]}$ is, in fact,
contracted with an $e_a^{\mu}$.  This means that we can compute the variation
of the original Lagrangian with respect to the tetrad in terms of its
variations
with respect the matter fields and their covariant derivatives:
\begin{eqnarray}
\label{initesource}
{{{\partial} {\cal L}^0} \over {{\partial} e_{\alpha}^a}} &=&
\case{1}{3!} \epsilon_{abcd} {\bf dx}^{\alpha} \wedge {\bf e}^b \wedge
{\bf e}^c \wedge {\bf e}^d L^0 \cr
&-& \sum_{n_i}
{{\partial {\cal L}^0} \over {\partial f^{(i)}{}_{[\mu_1 \mu_2 ...\mu_{n-1}
\alpha \mu_{n+1} ...]}}}
(e_{\mu_1}^{m_1} e_{\mu_2}^{m_2} ... e_{\mu_{n-1}}^{m_{n-1}}
e_{\mu_{n+1}}^{n+1} ...) (e_{m_1}^{\nu_1} e_{m_2}^{\nu_2} ...
e_{m_{n-1}}^{\nu_{n-1}} e_a^{\nu_n}
e_{m_{n+1}}^{\nu_{n+1}} ...) f^{(i)}{}_{[\nu_1 \nu_2
...]} \cr
&-&  \sum_{n,i}
{{\partial {\cal L}^0} \over {\partial {\cal D}_{[\beta}
f^{(i)}{}_{\mu_1 \mu_2 ...\mu_{n-1} \alpha \mu_{n+1} ...]}}}
(e_{\mu_1}^{m_1} e_{\mu_2}^{m_2} ... e_{\mu_{n-1}}^{m_{n-1}}
e_{\mu_{n+1}}^{m{n+1}} ...e_{\beta}^b) (e_{m_1}^{\nu_1} e_{m_2}^{\nu_2} ...
e_{m_{n-1}}^{\nu_{n-1}} e^{\nu_n}_a
e_{m_{n+1}}^{\nu_{n+1}} ... e_b^{\gamma})
{\cal D}_{[\gamma} f^{(i)}{}_{\nu_1 \nu_2
...]} \cr
&-& \sum_i {{\partial {\cal L}^0} \over {\partial {\cal D}_{[\alpha}
f^{(i)}{}_{\mu_1 \mu_2 ...]}}}  (e^{m_1}_{\mu_1} e^{m_2}_{\mu_2} ...)
(e_{m_1}^{\nu_1} e_{m_2}^{\nu_2} ...) e_a^{\gamma}
{\cal D}_{[\gamma} f^{(i)}{}_{\nu_1 \nu_2
...]} \cr
&=& \case{1}{3!} \epsilon_{abcd} {\bf dx}^{\alpha} \wedge {\bf e}^b \wedge
{\bf e}^c \wedge {\bf e}^d L^0 \cr
&-& \sum_{n,i} {{\partial {\cal L}^0} \over {\partial f^{(i)}{}_
{[\mu_1 \mu_2 ...\mu_{n-1} \alpha \mu_{n+1} ...]}}
}) f^{(i)}{}_{\mu_1 \mu_2 ...} e^{\mu_n}_a \cr
&-& \sum_{n,i} {{\partial {\cal L}^0} \over {\partial {\cal D}_{[\beta}
f^{(i)}{}_{\mu_1 \mu_2 ...\mu_{n-1} \alpha \mu_{n+1}]}}}
{\cal D}_{\beta} f^{(i)}{}_
{\mu_1 \mu_2 ...} e^{\mu_n}_a  \cr
&-& \sum_i {{\partial {\cal L}^0} \over {\partial {\cal D}_{\alpha}
f^{(i)}{}_{\mu_1 \mu_2 ...}}}  {\cal D}_{\beta} f^{(i)}{}_
{\mu_1 \mu_2 ...} e^{\beta}_a
\end{eqnarray}

If we now inspect Eq. \ref{esource} term by term, we see that these
terms become
exactly the terms in Eq. \ref{initesource} when the equations \ref{defs} (or,
more directly, Eq. \ref{varfields} and \ref{varderivatives}) are
used to substitute for the various Lagrange multipliers.
It follows that when the matter equations of motion
hold, the variation of our new action with respect to the
tetrad is equal to the corresponding variation of the old action.

\section{Generalization}
\label{gen}

A generalization of the algorithm presented in section \ref{alg}
allows the original action to contain
arbitrary matter fields, but has the disadvantage that
it introduces contravariant tensor fields.  However, this
algorithm introduce less contravariant fields than previous methods
\cite{others,others2} and, despite the presence of the contravariant fields,
the
resulting actions are still invariant under a sort of
pull back in certain cases.  The algorithm is given below and is followed by a
short
discussion of the invariance.
We do not present a separate proof that the resulting actions
are equivalent to the original ones since such a proof is very much the
same as the one given in section \ref{proof}.

\begin{enumerate}

\item [Step 1)]  For each tensor density $T^{[w](i)}{}_{\mu \nu
...}$ of weight $w \neq 0$, introduce a new tensor
field $T^{(i)}{}_{\mu \nu ...}$ and use it replace that density by
performing the
substitution:
\begin{equation}
T^{[w](i)}{}_{\mu \nu ...} \rightarrow [det(e)]^w T^{(i)}{}_{\mu \nu ...}
\end{equation}
where $det(e)$ is the determinant of the tetrad $e$.

\item [Step 2)]  Insert sufficient inverse tetrads to
write all tensor fields in terms of their covariant
components:
\begin{equation}
T^{(i)\alpha}{}_{\mu \nu ...} \rightarrow e^{\alpha}_a \eta^{ab}
e^{\beta}_b T^{(i)}{}_{\beta \mu \nu ...}
\end{equation}

\item [Step 3)]  Insert enough tetrads to write all
tensor fields in terms of their tetrad components.
More specifically, for each tensor field $T^{(i)}{}_{\mu \nu
...}$ (including those introduced in Step 1) introduce a
set of scalar fields $T^{(i)}{}_{m n ...}$ with the same number
of internal indices as the rank of the original tensor
field and with the same symmetries.  Then perform the
substitution:
\begin{equation}
T^{(i)}{}_{\mu \nu ...} \rightarrow e^m_{\mu} e^n_{\nu} ...
T^{(i)}_{}{mn ...}
\end{equation}

\item [Step 4)]  Arrange the terms now present in the
``action" so that the covariant derivatives act only on
spacetime scalars, though these may have an arbitrary
structure of internal indices.

\item [Step 5)]  For each tensor field $T^{(i)}{}_{\mu \nu ...}$
either present in the original action or introduced in
Step 1, introduce another collection of spacetime
scalars $T^{(i)}{}_{mn ...,a}$ labeled by one more internal
index than the rank of the tensor field.  Again, this
comma is {\em only} a grouping symbol and does not denote
{\em any} kind of differentiation.  Now, introduce these
new fields into the action by using them to replace the
covariant derivatives of the fields $T^{(i)}{}_{mn ...}$
according to the rule:
\begin{equation}
{\cal D}_{\alpha}T^{(i)}{}_{mn ...} = e^a_{\alpha} e^m_{\mu}
e^n_{\nu} T^{(i)}{}_{mn ...,a}
\end{equation}
Again, no covariant derivatives remain in the
Lagrangian after this substitution has been performed.

\item [Step 6)]  Replace any spacetime Levi-Civita
densities with the corresponding internal densities:
\begin{equation}
\epsilon_{\alpha \beta \gamma \delta} \rightarrow
\epsilon_{abcd}
e^a_{\alpha}e^b_{\beta}e^c_{\gamma}e^d_{\delta}
\ \text{and} \ \epsilon^{\alpha \beta \gamma \delta}
\rightarrow \epsilon^{abcd}
e_a^{\alpha}e_b^{\beta}e_c^{\gamma}e_d^{\delta}
\end{equation}
and replace $det(e) d^4x$ with $\case {1}{4!} e^a \wedge
e^b  \wedge e^c \wedge e^d$.

\item [Step 7)]  Formally cancel all remaining positive and
negative powers of $det(e)$ and all contracted tetrads and
inverse tetrads:
\begin{equation}
e_m^{\mu} e_{\mu}^n \rightarrow \delta_m^n \ \text{and} \
e_{\mu}^m e_m^{\nu} \rightarrow \delta_{\mu}^{\nu}
\end{equation}
Again, the tetrad no
longer appears in the action except though the four-form:
$\case {1}{4!} e^a \wedge e^b  \wedge e^c \wedge e^d$.

\item  [Step 8)]  For each tensor field $T^{(i)}_{[\mu \nu ...]}$
in the original action or introduced in Step 1,
introduce a set of four-form fields
$\kappa^{(i)}{}_{mn...}$ with a number of internal indices
equal to the rank of the tensor field and a set of
three-form fields $\lambda^{(i)}{}_{mn...}$ also with a
number of internal indices given by the rank of the
tensor field.  Now, use these new fields as Lagrange
multipliers and add to the above Lagrange density the
constraint terms:
\begin{equation}
\kappa^{(i)}{}_{mn...} [T^{(i)\mu \nu ...} e_{\mu}^m
e_{\nu}^n ... - T^{(i)mn ...}]
\end{equation}
and
\begin{equation}
\label{diff}
\lambda^{(i)}{}_{mn...} \wedge [{\cal D} (T^{(i)\mu \nu ...}
e_{\mu}^m e_{\nu}^n ...) - T^{(i)mn ...}_{,a} {\bf e}^a]
\end{equation}
where the ${\cal D}$ with no subscript is an external
derivative operator that creates a one-form from the
spacetime scalar field on which it acts.  The internal indices
on the matter tensor field are raised using the appropriate internal metric.
\end{enumerate}

The resulting action contains the contravariant tensor
fields $T^{(i)\mu_1 \mu_2 ...}$, but only when contracted with tetrads.
As a result, any transformation that leaves these contractions invariant
is a gauge transformation.  This result can be used to show that even
these actions are invariant under a sort of pull back, where the new
``pulled back"
values $T'{}^{(i)\mu_1 \mu_2 ...}$ of this contravariant tensor field
are chosen to be any values such
that their contraction with the pulled back tetrad  $e'{}^{m}_{\mu}$
are the same as the old
contractions:
\begin{equation}
T'{}^{(i)\mu_1 \mu_2 ...} e'{}^{m_1}_{\mu_1} e'{}^{m_2}_{\mu_2} ...
= T{}^{(i)\mu_1 \mu_2 ...} e^{m_1}_{\mu_1} e^{m_2}_{\mu_2} ...
\end{equation}
The action is then invariant under any ``pull back"
of degree one for which such new components $T'{}^{(i)\mu_1 \mu_2 ...}$
of $T^{(i)\mu_1 \mu_2 ...}$ exist.

Since these contravariant fields appear only
in constraints and only in contractions with tetrads it might seem
that they could be
eliminated entirely even from this more general formalism.  This is not
easy to do and simple attempts to remove them do not work.  For example,
if the contravariant fields are replaced by covariant tensor fields,
then the resulting covariant indices must be contracted with some other
fields.
This requires either the introduction of inverse tetrads to supply the needed
contravariant indices, the introduction of contravariant indices on the
Lagrange
multipliers $\lambda$ and $\kappa$, or the contraction of these covariant
indices with the forms $dx^{\mu} \wedge dx^{\nu} \wedge ...$ which is
essentially what was done in section \ref{alg} for the class of matter
fields with the appropriate symmetries.  Another option is to remove
the contravariant fields entirely and to replace them in the
differential constraint by their tetrad components.  While
the resulting matter dynamics is correct, such an action contains
tetrads only through the volume element and through the single tetrad in
each differential constraint \ref{diff}.
But this is completely the wrong coupling to
the gravitational field, and we are forced to keep the contravariant fields.

\section{Discussion}

We have seen that it is possible to write actions for certain kinds of
matter fields that remain well-defined when the tetrad is degenerate
and contain only fields that can be pulled back.  In fact, these
actions are invariant under the pull back of these fields through any
map of degree one.  It would thus seem that we should identify a
set of fields and its pull back as gauge related, at least if the pull back
map is continuously connected to the identity map.

This point was raised in \cite{Gary} in the purely gravitational context
and is not significantly different here.  There, this comment was followed
by a discussion of how the gauge group might be enlarged so that this is
so.  We would simply like to make the comment that while it is nice
to identify a gauge group, such an identification is not strictly necessary
for the identification and investigation of gauge {\em orbits}.  Since
gauge transformations are defined in an infinitesimal form (see, for example
\cite{Bryce}) with no stipulation that they can be integrated in such a
way that the finite transformations form a group, we can use the
infinitesimal transformations to define the gauge orbits.  That is,
two field histories are gauge related if and only if they can be
continuously connected by a set of infinitesimal transformations.  This
relation is necessarily an equivalence relation and so can be used to define
the physical equivalence classes without reference to any gauge group.

Having said this, we would like to make one further suggestion with
regard to these physical equivalence classes.  Whenever the ``gauge
group" is not connected, there is always the question of whether
to identify field histories that are related by large gauge transformations;
that is, by gauge transformations not continuously connected to the identity.
A similar issue arises here:  only pull backs via maps of degree one
can be continuously connected to the identity so that only such pull backs
necessarily impose physical equivalence between field histories.  However,
we would like to suggest that histories related by
pull backs through maps of any degree other than
zero might also be considered physically equivalent.

We suggest this
despite the fact that such transformations do not in fact leave the
action invariant (as large gauge transformations would do), but multiply it
by the degree of the map.  Nevertheless, a number of arguments could be
made that this is a physically reasonable thing to do, based on the
observational indistinguishability of such histories.  These arguments
discuss measurements performed {\em in} the two field histories.
We stress the word {\em  in} because for such arguments it is
important that the fields that define the laboratory and measuring apparatus
also be a part of our description and that they too be pulled back from
one history to the other.  In such a setting, any field configuration that
describes an experiment and result in one spacetime is pulled back to
describe the same experiment and the same result in the other spacetime.
While there are, at least in principle, global measurements that could
distinguish between the two spacetimes, the experimental apparatus
required
are not related by pull backs
and such measurements
are never performed by experimentalists living in such spacetimes.

Potential problems with this suggestion are the possibilities of
excessively large equivalence classes resulting from such identifications
and of complicated topologies on the space of such classes.  For example,
the space of solutions of the equations of motion on a cylinder would not
be disconnected from the space of solutions on two cylinders, but would
be connected through the two cylinder pull backs of one cylinder solutions.
This might also make any sum over histories formulation that allows topology
change even more complicated since the action would no longer be a continuous
(or even well-defined) functional on the space of gauge
equivalence classes.  I will not comment further on the
possible implications for ``topology change"
since a number of interpretations are possible without a definite structure
in which to work.  Note, however, that with this definition of
equivalence the classical topology changing solutions
described in \cite{Gary} would be considered equivalent to solutions
that do not change topology.  It is not clear whether these solutions can
be generalized in a straightforward way so that they are no longer the
pull back of some solution that does not change topology.

\acknowledgements
This work was partially supported by NSF grant
PHY-9005790 and by research funds provided by Syracuse
University.


\begin{references}
\bibitem{Marolf}[*] Email: marolf@suhep.ph.syr.edu
Address after August 1, 1993: Department of Physics, The Pennsylvania
State University, Univeristy Park, PA 16802

\bibitem{Witten} E Witten 1988 {\it Nucl. Phys.} {\bf B311} 46-78

\bibitem{Ash} A Ashtekar 1990 {\it Lectures on Nonperturbative Canonial
Gravity} (World Scientific: Singapore)

\bibitem{Gary} G Horowitz 1991 {\it Class. Quantum Grav.}
{\bf 8} 587-601

\bibitem{Palatini} A Palatini 1919 {\it Rend. Circ. Mat. Palermo} {\bf 43}
203

\bibitem{cones} G Hayward and J Louko 1990 {\it Phys. Rev. D} {\bf 42}
4232-4041

\bibitem{others} A Tseytlin 1982 {\it J. Phys. A: Math. Gen.} {\bf 15}
L105

\bibitem{others2} R Capovilla, J Dell, T Jacobsen, and L Mason 1991
{\it Class. Quant. Grav.} {\bf 8} 41

\bibitem{Bryce}  B DeWitt 1984
in {\it Relativity, Groups, and Topology: Les Houches
1983} Ed. by B DeWitt and R Stora (North-Holland: New York)


\end{references}
\end{document}